\documentclass[
 amsmath,amssymb,
 aps,
]{revtex4-2}

\pdfoutput=1

\usepackage[colorlinks,urlcolor=blue,citecolor=blue]{hyperref}
\def\hhref#1{\href{http://arxiv.org/abs/#1}{arXiv:#1}} 

\usepackage{graphicx}
\usepackage{bm}

\begin{document}

\title{Resurgence of the Tilted Cusp Anomalous Dimension}

\author{Gerald V. Dunne}
 \affiliation{Physics Department,  University of Connecticut, Storrs CT 06269, USA.}


\begin{abstract}
We use resurgent extrapolation and continuation methods to extract detailed analytic information about the tilted cusp anomalous dimension solely from its weak coupling and strong coupling expansions. This enables accurate and smooth interpolation between the weak and strong coupling limits, and identifies the relevant singularities governing the finite radius of convergence of the weak coupling expansion and the asymptotic nature of the strong coupling expansion. The input data is purely perturbative, generated from the BES equations, and these resurgent methods extract accurate non-perturbative information which matches the underlying physical structure.

\end{abstract}

\maketitle


\centerline{\it Dedicated to the memory of Stanley Deser}
\medskip

\section{\label{sec:level1}Introduction}

It has recently been shown that in a certain kinematical limit the maximally-helicity-violating amplitude for the scattering of six gluons in $\mathcal N=4$ supersymmetric Yang-Mills theory involves a function, the ``tilted cusp'' \cite{Basso:2020xts}, which is a deformation of the conventional cusp anomalous dimension $\Gamma_{\rm cusp}$. 
The tilted cusp anomalous dimension is defined in terms of the Beisert-Eden-Staudacher (BES) kernel \cite{Beisert:2006ez,Basso:2007wd,Basso:2009gh}
\begin{eqnarray}
K_{ij}=2j (-1)^{j(i+1)} \int_0^\infty \frac{dt}{t} \frac{J_i(2g t) J_j(2 g t)}{e^t-1}
\label{eq:besk}
\end{eqnarray}
Then one defines \cite{Basso:2020xts}
\begin{eqnarray}
    \Gamma_a=4g^2\left[\frac{1}{\mathbb{I}+\mathbb K(a)}\right]_{11}
    \label{eq:gamma-a1}
\end{eqnarray}
in terms of the $(11)$ component of the inverse of the infinite dimensional matrix $\left[\mathbb{I}+\mathbb K(a)\right]$ where:
\begin{eqnarray}
    \mathbb K(a):=2\, \cos(a\,\pi)\left[\begin{matrix}
        \cos(a\, \pi) \mathbb K_{oo} & \sin(a\, \pi) \mathbb K_{oe} \\
        \sin(a\, \pi) \mathbb K_{eo} & \cos(a\, \pi) \mathbb K_{ee}
    \end{matrix} \right]
    \label{eq:ka}
\end{eqnarray}
This matrix is sub-blocked with respect to odd (o) and even (e) indices of $K_{ij}$ in the kernel \eqref{eq:besk}.
The tilted cusp is parametrized by the tilt angle parameter 
\begin{eqnarray}
    a=\frac{\alpha}{\pi}\qquad, \qquad a\in \left[0,\frac{1}{2}\right]
    \label{eq:a}
\end{eqnarray}
We also recall that $g$ is related to the 't Hooft coupling 
\begin{eqnarray}
    \lambda:=g_{\rm YM}^2 N_c=(4\pi g)^2
    \label{eq:thooft}
\end{eqnarray}
The usual cusp anomalous dimension corresponds to $a=\frac{1}{4}$, the hexagon to $a=\frac{1}{3}$, and the octagon to $a=0$ \cite{Basso:2020xts}. The octagon case ($a=0$) and the $a=\frac{1}{2}$ case are known in closed-form
\begin{eqnarray}
    \Gamma_{0}&=&\frac{2}{\pi^2} \ln\cosh(2\pi g)
    \label{eq:oct}
    \\
    \Gamma_{\frac{1}{2}}&=&4g^2
    \label{eq:ahalf}
\end{eqnarray}
but in other cases we rely on expansions at weak and strong coupling, or numerical evaluations. A general approach to the tilted cusp is to explore its deep relation to the Tracy-Widom distribution \cite{Bajnok:2024epf,Bajnok:2024ymr}.

In this paper we apply methods developed for resurgent asymptotic expansions \cite{Costin:2020hwg,Costin:2020pcj,Costin:2021bay} to extract singularity information about the tilted cusp from its convergent weak coupling expansion, and non-perturbative information from the asymptotic strong coupling expansion. In order to demonstrate that this analytic information is encoded in the perturbative expansions, no explicit use is made of \eqref{eq:besk}-\eqref{eq:ka}. This analysis builds on, and extends, previous analyses of the resurgent properties of the usual cusp anomalous dimension $\Gamma_{\rm cusp}$ \cite{Aniceto:2015rua,Dorigoni:2015dha}.
 
\section{Extrapolating the Weak Coupling Expansion}
\label{sec:weak-exrapolation}

\subsection{Weak Coupling Expansion}
\label{sec:weak}
The weak coupling expansion is conveniently expressed in the normalized form:
\begin{eqnarray}
    \frac{\Gamma_a}{4g^2}:=\sum_{n=0}^\infty b_n(a) g^{2n} 
    \label{eq:gamma-weak}
\end{eqnarray}
with expansion coefficients $b_n(a)$ that depend on the tilt parameter $a$ (except for $b_0(a)= 1$, for all $a$). The first few terms are \cite{Basso:2020xts}:
\begin{eqnarray}
   \frac{\Gamma_a}{4g^2}=1-4c^2 \zeta_2 g^2+8c^2(3+5c^2)\zeta_4 g^4-8c^2\left[(25+42c^2+35c^4)\zeta_6+4s^2\zeta_3^2\right]g^6+\dots
   \label{eq:gamma-weak-a}
\end{eqnarray}
where $c\equiv \cos(\pi a)$ and $s\equiv \sin(\pi a)$.
For the cusp, hexagon and octagon the first few terms are:
\begin{eqnarray}
    \frac{\Gamma_{\rm cusp}}{4g^2}=1-\frac{\pi ^2 g^2}{3} +\frac{11 \pi^4 g^4}{45}+g^6 \left(-8 \zeta (3)^2-\frac{73 \pi ^6}{315}\right) +g^8 \left(\frac{16 \pi ^2 \zeta (3)^2}{3}+160 \zeta (3) \zeta (5)+\frac{3548 \pi
   ^8}{14175}\right) +\dots  
   \label{eq:cusp-weak}
\end{eqnarray}

\begin{eqnarray}
    \frac{\Gamma_{\rm hex}}{4g^2}=
    1 -\frac{\pi ^2 g^2}{6} + \frac{17 \pi
   ^4 g^4}{180}  +g^6 \left(-6 \zeta (3)^2-\frac{67 \pi ^6}{840}\right)
    +g^8 \left(2 \pi ^2 \zeta (3)^2+120 \zeta (3) \zeta (5)+\frac{18287 \pi
   ^8}{226800}\right)+\dots
   \label{eq:hex-weak}
   \end{eqnarray}

   \begin{eqnarray}
       \frac{\Gamma_{\rm oct}}{4g^2}=
       1-\frac{2 \pi ^2 g^2}{3}+\frac{32 \pi ^4 g^4}{45}-\frac{272 \pi ^6 g^6}{315}+\frac{15872
   \pi ^8 g^8}{14175}-\frac{707584 \pi ^{10} g^{10}}{467775}+ \dots
   \label{eq:oct-weak}
   \end{eqnarray}
Notice that the octagon case is special (given the closed-form in \eqref{eq:oct}), with all the odd-zeta terms being absent.

\begin{figure}[h!]
\centerline{\includegraphics[scale=.6]{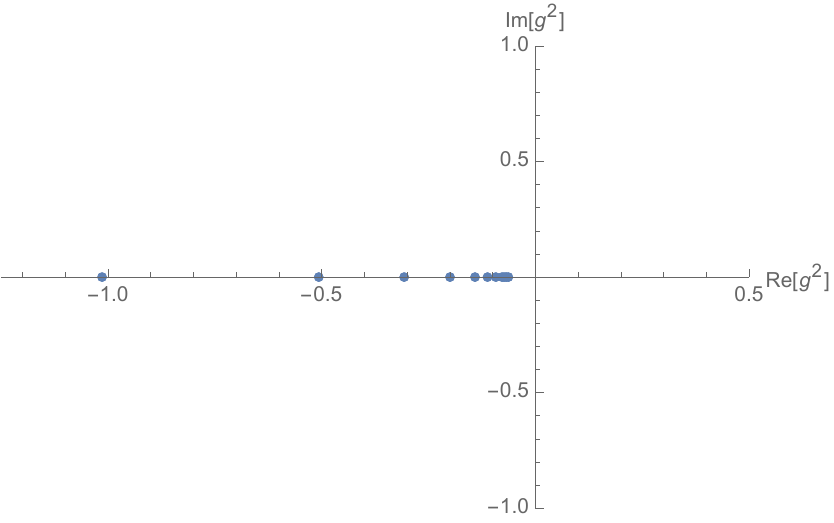}
\qquad
\includegraphics[scale=.6]{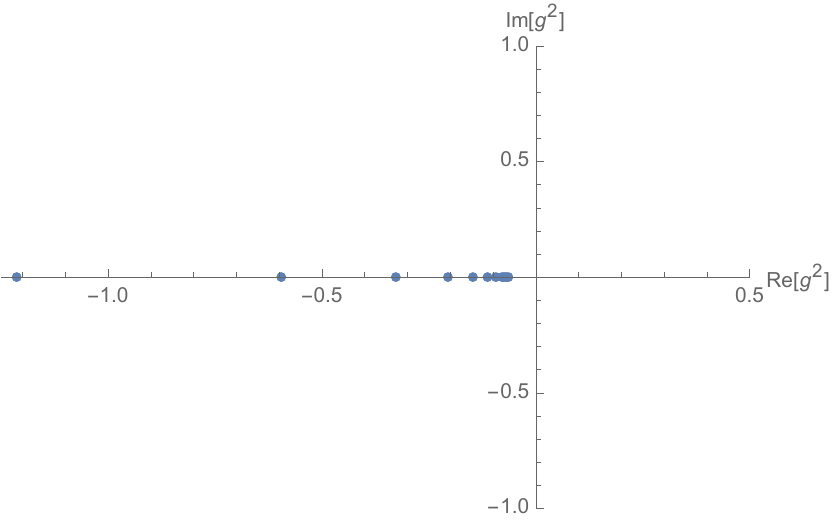}}
\caption{Pad\'e poles for the weak coupling expansions for the cusp [left] and hex [right], in the complex $g^2$ plane. In each case the poles accumulate to the branch point at $g^2=-\frac{1}{16}$.}
\label{fig:weak-pade-poles}
\end{figure}

The weak coupling expansion \eqref{eq:gamma-weak} is a convergent expansion in $g^2$, with radius of convergence $\frac{1}{16}$, for all $0\leq a<\frac{1}{2}$, due to a singularity at $g^2=-\frac{1}{16}$. This is straightforwardly identified numerically by Pad\'e approximants: see Figure \ref{fig:weak-pade-poles}.
And since there are no singularities in the direction of $g^2$ positive, the Pad\'e approximants also provide a simple and accurate extrapolation of the weak coupling expansion to larger positive values of $g^2$. 
\begin{figure}[h!]
\centerline{\includegraphics[scale=.5]{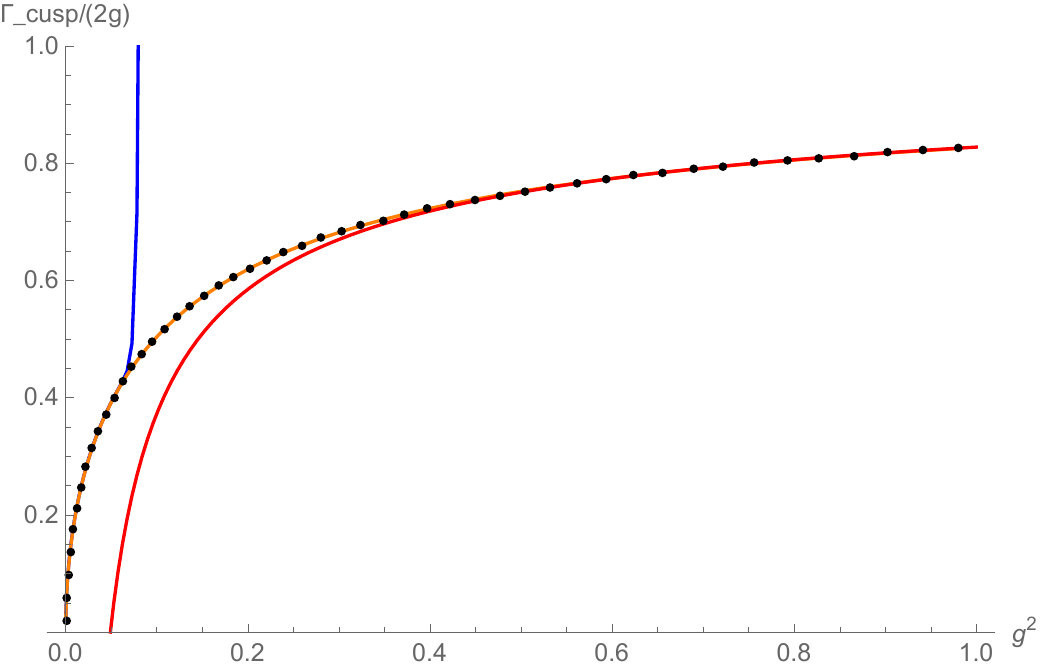}
\includegraphics[scale=.5]{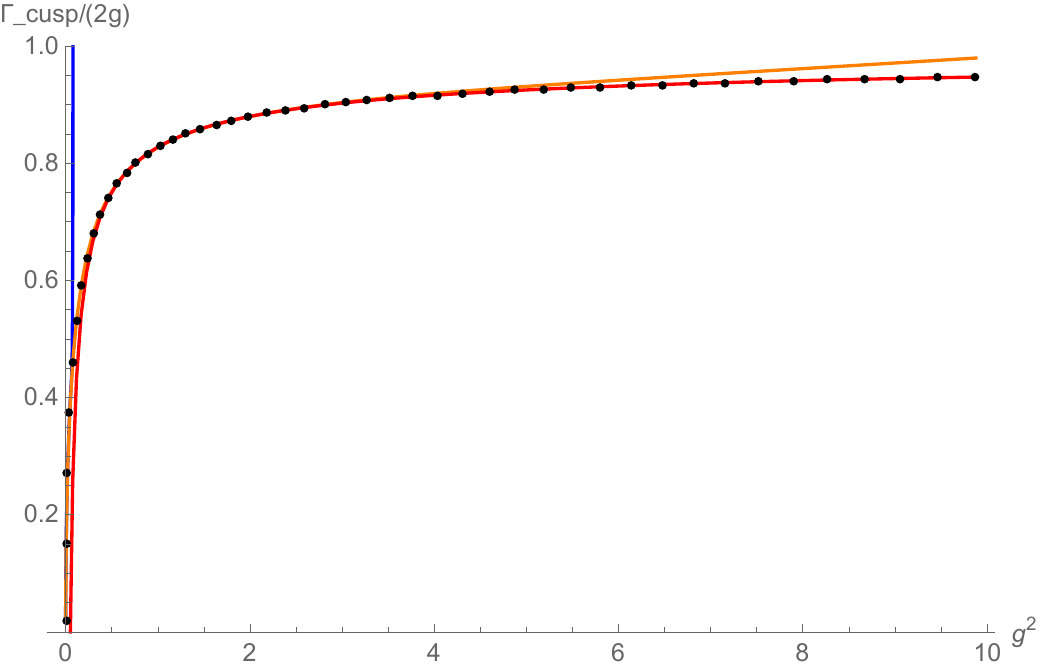}}
\caption{Extrapolation of the weak coupling expansion of the cusp ($a=\frac{1}{4}$) based on 24 terms of the weak coupling expansion. The left-hand plot shows the extrapolation out to $g^2=1$, and the right-hand plot extends out to $g^2=10$. 
The blue curve is the 24-term weak coupling series expansion, whose breakdown at the radius of convergence, $g^2=\frac{1}{16}$, can be clearly seen. The red curves plot the first 3 terms of the (divergent) strong coupling expansion \eqref{eq:cusp-strong}.
The orange curves are Pad\'e approximants of the 24-term weak coupling series.  The black dots show the Pad\'e-Conformal approximation, which extrapolates much further towards strong coupling.}
\label{fig:cusp-weak}
\end{figure}
\begin{figure}[h!]
\centerline{\includegraphics[scale=.5]{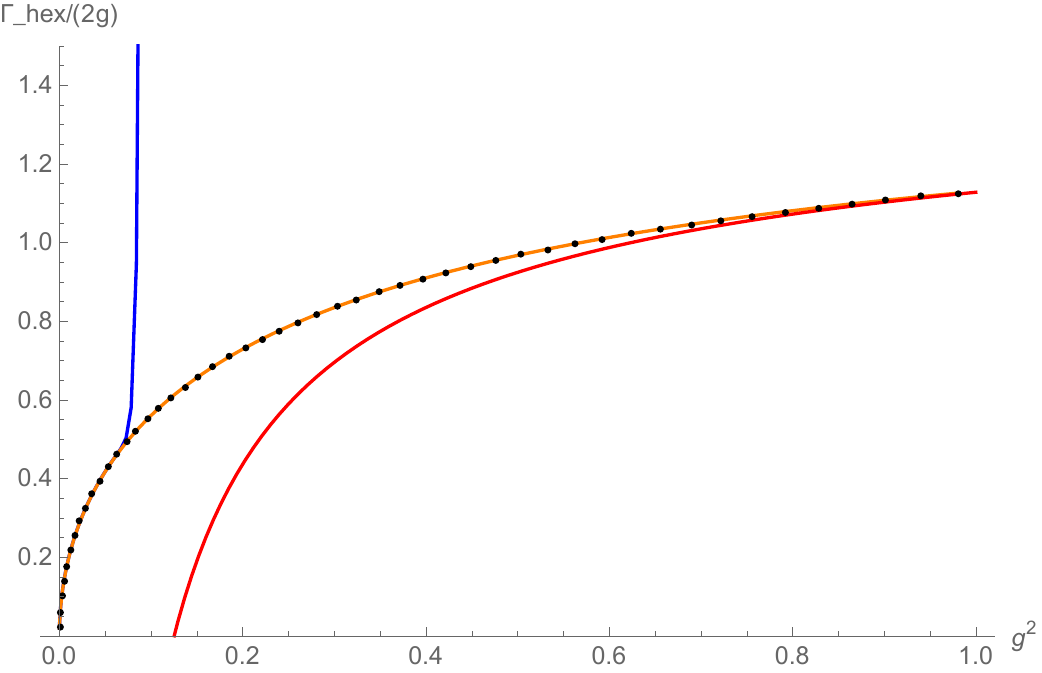}
\includegraphics[scale=.5]{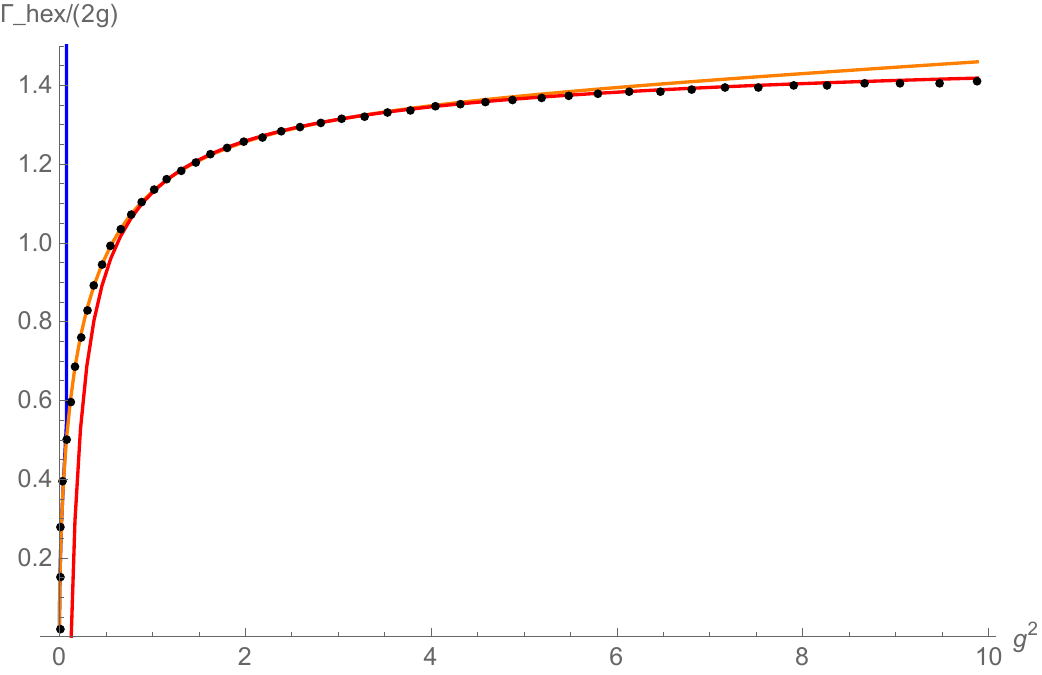}}
\caption{Extrapolation of the weak coupling expansion of the hex ($a=\frac{1}{3}$) based on 24 terms of the weak coupling expansion. The left-hand plot shows the extrapolation out to $g^2=1$, and the right-hand plot extends out to $g^2=10$. The blue curve is the 24-term weak coupling series expansion, whose breakdown at the radius of convergence, $g^2=\frac{1}{16}$, can be clearly seen. The red curves plot the first 3 terms of the (divergent) strong coupling expansion \eqref{eq:hex-strong}.
The orange curves are Pad\'e approximants of the 24-term weak coupling series.  The black dots show the Pad\'e-Conformal approximation, which extrapolates much further towards strong coupling.}
\label{fig:hex-weak}
\end{figure}
In Figures \ref{fig:cusp-weak} and \ref{fig:hex-weak}, the blue curve shows the weak coupling expansion truncated at order $g^{50}$, for the cusp anomalous dimension and for the hex anomalous dimension, respectively.  In each figure, the left-hand plot shows the extrapolation from $g^2=0$ out to $g^2=1$, while the right-hand plot extends from $g^2=0$ out to $g^2=10$. 
The finite radius of convergence is clearly seen, as the blue curve diverges at $g^2=\frac{1}{16}$ in each case. The orange curves in these plots show the Pad\'e approximant, while the red curve shows the first 3 terms of the strong-coupling expansions in \eqref{eq:cusp-strong} and \eqref{eq:hex-strong}.
Observe that even a simple Pad\'e approximant extrapolates quite accurately beyond the radius of convergence, although it begins to deviate from the strong-coupling behavior around $g^2=5$.

The black dots in Figures \ref{fig:cusp-weak} and 
\ref{fig:hex-weak} show the result of a conformal-Pad\'e extrapolation, which extends more accurately into the large $g^2$ region. The conformal-Pad\'e extrapolation is simple to generate and is provably more accurate than Pad\'e  
\cite{Costin:2020hwg,Costin:2020pcj,Costin:2021bay}. The general procedure is: 
\begin{enumerate}
    \item 
Make a conformal map of the cut complex $g^2$ plane to the interior of the unit disk in the $z$ plane
\begin{eqnarray}
    16g^2=\frac{4z}{(1-z)^2} \qquad \longleftrightarrow \qquad
z=\frac{\sqrt{1+16g^2}-1}{\sqrt{1+16g^2}+1}
\label{eq:weak-conformal-map}
\end{eqnarray}
and re-expand about $z=0$ to the same number of terms as the original expansion (this is provably optimal \cite{Costin:2020pcj}).
\item Make a Pad\'e approximant inside the unit disk in the $z$ plane.
\item Finally, map this Pad\'e approximant back to the $g^2$ plane via the inverse conformal map.
\end{enumerate}
The black dots in Figures \ref{fig:cusp-weak} and 
\ref{fig:hex-weak} illustrate the improvement out to large $g^2$.

\begin{figure}[h!]
\centerline{\includegraphics[scale=.6]{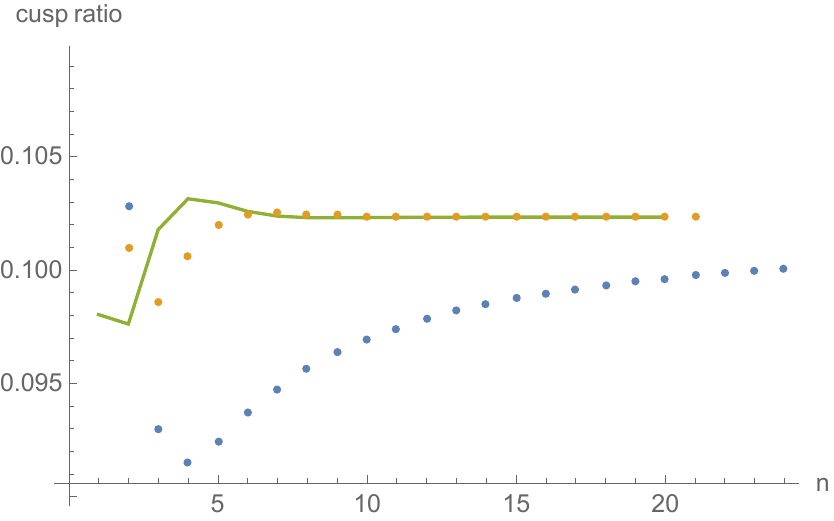}
\qquad
\includegraphics[scale=.6]{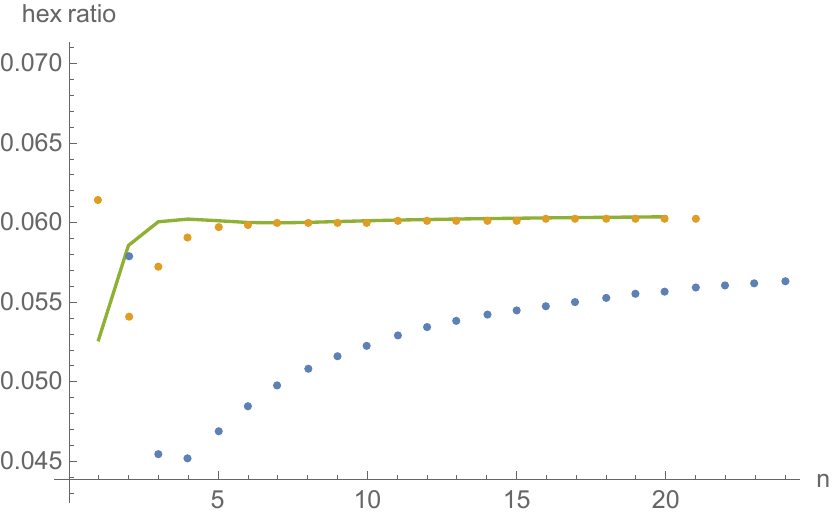}}
\caption{The Darboux test ratios for the cusp [left] and hex [right], showing the ratio of the weak-coupling coefficients to the form in \eqref{eq:darboux}. The blue dots show the raw ratio; the orange dots and green curve show the 4th and 5th order Richardson acceleration, respectively.}
\label{fig:darboux}
\end{figure}

The coefficients of the weak coupling expansion also encode information about the {\it nature} of the singularity at 
$g^2=-\frac{1}{16}$. This follows from Darboux's theorem \cite{henrici}, which states that the large order behavior of the coefficients of the expansion about the origin encodes information about the expansion near the singularities. For example, numerical analysis from 24 terms of the weak coupling expansions suggests the {\it leading} large order behavior (for $0<a<\frac{1}{2}$):
\begin{eqnarray}
    b_n(a)\sim \mathcal C\, (-16)^n \begin{pmatrix}n-(1+2a) \\ n\end{pmatrix}
    \left(1+ O\left(\frac{1}{n}\right)\right) +\dots
    \label{eq:darboux}
\end{eqnarray}
See Figure \ref{fig:darboux}, which plots the ratio of the weak-coupling coefficients to the leading behavior in \eqref{eq:darboux}. The blue dots are the raw ratio, while the orange dots and the green curve show the results of fourth and fifth order Richardson acceleration \cite{bender} of the ratio. The factor $(-16)^n$ in \eqref{eq:darboux} is simply the statement that the singularity closest to the origin is at $g^2=-\frac{1}{16}$, for all $a$. But the $a$ dependence in the binomial factor indicates that the exponent of the singularity depends on the tilt parameter $a$:
\begin{eqnarray}
    \frac{\Gamma_a}{4g^2}\sim  \mathcal C \left(1+16g^2\right)^{2a}\left[
1+O\left(1+16g^2\right)\right] + \dots 
\label{eq:darboux-a}
\end{eqnarray}
where $\dots$ indicates terms regular at $g^2=-\frac{1}{16}$. The constant $\mathcal C$ depends on $a$ in a non-trivial way. Note that expression \eqref{eq:darboux-a} is also consistent with the fact that the limiting values for $a\to 0$ and $a\to \frac{1}{2}$ are special. The singularities for $a=0$ are logarithmic 
\begin{eqnarray}
    b_n(0)= \frac{8}{\pi^2} \frac{(-16)^n}{(n+1)}(1-2^{-2n-2})\zeta_{2n+2}  
\qquad \Rightarrow \qquad \Gamma_0 =  -\frac{2}{\pi^2} \log\left(\Gamma\left(\frac{1}{2}-2i g\right) \Gamma\left(\frac{1}{2}+2i g\right)/\pi\right) 
\label{eq:darboux-0}
\end{eqnarray}
And for $a=\frac{1}{2}$ the tilted cusp becomes regular: $\Gamma_{\frac{1}{2}}=4g^2$.

\section{Extrapolating the Strong Coupling Expansion}
\label{sec:strong-extrapolation}

\subsection{The Strong Coupling Expansion}
\label{sec:strong}

We write the strong coupling expansion in the convenient normalized form:
\begin{eqnarray}
    \Gamma_a\sim \frac{2 a}{\pi\sin(2\pi a)}\, \xi \sum_{n=0}^\infty \frac{c_n(a)}{\xi^n} 
\label{eq:gamma-strong}
\end{eqnarray}
in terms of the shifted coupling (see \cite{Basso:2007wd,Basso:2020xts}; and recall $\lambda$ is the 't Hooft coupling \eqref{eq:thooft})
\begin{eqnarray}
    \xi:=4\pi g-\frac{s_1(a)}{2}=\sqrt{\lambda}-\frac{s_1(a)}{2}
    \label{eq:xi}
\end{eqnarray}
With this normalization, $c_0(a)=1$ and $c_1(a)=0$, for all $0\leq a< \frac{1}{2}$. Here the $s_k(a)$ are functions of the tilt parameter $a$ \cite{Basso:2020xts}: 
\begin{eqnarray}
s_{k+1}(a)=\left[\psi_k(1)-\psi_k\left(\frac{1}{2}+a\right)\right]    +(-1)^k \left[\psi_k(1)-\psi_k\left(\frac{1}{2}-a\right)\right] 
\label{eq:sk}
\end{eqnarray}
where $\psi_k(z):=\left(\frac{d}{dz}\right)^{k+1}\ln\Gamma(z)$. Note that $s_k(a)$ develops an order $k$ pole at $a=\frac{1}{2}$, so the limit $a\to \frac{1}{2}$ is most naturally studied in a double-scaling limit \cite{unpub,Bajnok:2024bqr}.

The first few strong-coupling expansion coefficients are \cite{Basso:2020xts}:
\begin{eqnarray}
\Gamma_a= \frac{2 a}{\pi\sin(2\pi a)}\, \xi \left[1-\frac{a s_2}{4\xi^2}-\frac{a^2 s_3}{8\xi^3} -\frac{a(12 a s_2^2+(1+5a^2)s_4}{96\xi^4}+\dots\right]
    \label{eq:strong-a}
\end{eqnarray}
For the cusp $(a=\frac{1}{4})$, hexagon $(a=\frac{1}{3})$ and octagon $(a=0)$ cases we have (here ${\bf K}$ is the Catalan constant):
\begin{eqnarray}
    \Gamma_{\rm cusp}&\sim& \frac{\xi}{2 \pi} 
    \left(1-\frac{{\bf K}}{\xi^2}-\frac{27\zeta_3}{32\xi^3} +\dots\right)
   \label{eq:cusp-strong}
   \\
  \Gamma_{\rm hex}&\sim& \frac{4\xi }{3\sqrt{3}\, \pi}\left(1-\frac{\psi_1\left(\frac{1}{6}\right)-
  \psi_1\left(\frac{5}{6}\right)}{12\xi^2}-\frac{5\zeta_3}{\xi^3} +\dots\right)
   \label{eq:hex-strong}
\\
 \Gamma_{\rm oct}
&=& \frac{\xi}{\pi^2} +\frac{2}{\pi^2} \sum_{k=1}^\infty \frac{(-1)^{k+1}}{k} \, \frac{e^{-k\, \xi}}{4^k}
\label{eq:gamma0}
\end{eqnarray}
The strong coupling expansion \eqref{eq:gamma-strong} is generically an asymptotic expansion, with factorially growing coefficients, as discussed below. However, for the octagon case ($a=0$), the perturbative part of the strong coupling expansion truncates after just 1 term, but includes an explicit non-perturbative transseries part. This is an example of the "Cheshire Cat" phenomenon, in which a high degree of symmetry can truncate a generically asymptotic perturbative expansion but still leave a trace of that divergence in additional non-perturbative terms \cite{gw,Wadia:1980cp,dk,Dunne:2016jsr,Kozcaz:2016wvy}.

Accurate extrapolation of the asymptotic strong coupling expansion from large $g^2$ down to small $g^2$ (equivalently, large $\xi$ to small $\xi$) can be achieved by applying optimal extrapolation methods to the Borel transform function, which has a finite radius of convergence, rather than to the tilted cusp function itself. See Section \ref{sec:borel}. The precision of the extrapolation of the tilted cusp function is determined by the precision with which the analytic structure of the Borel transform is known, especially near its singularities \cite{Costin:2020hwg,Costin:2020pcj}.

\subsection{Large Order Growth of Strong Coupling Expansion Coefficients}
\label{sec:growth}

Straightforward numerical tests indicate that the coefficients of the perturbative strong-coupling expansion \eqref{eq:gamma-strong} have large-order behavior of the canonical factorial-over-power form (for $0<a <\frac{1}{2}$):
\begin{eqnarray}
c_n (a)\sim \mathcal S\,  \frac{\Gamma(n-\beta)}{A^n}\left(1+A \frac{b}{(n-\beta-1)} 
+ O \left(\frac{1}{n^2}\right)\right)+\dots 
\label{eq:cn}
\end{eqnarray}
\begin{figure}[h!]
\includegraphics[scale=0.6]{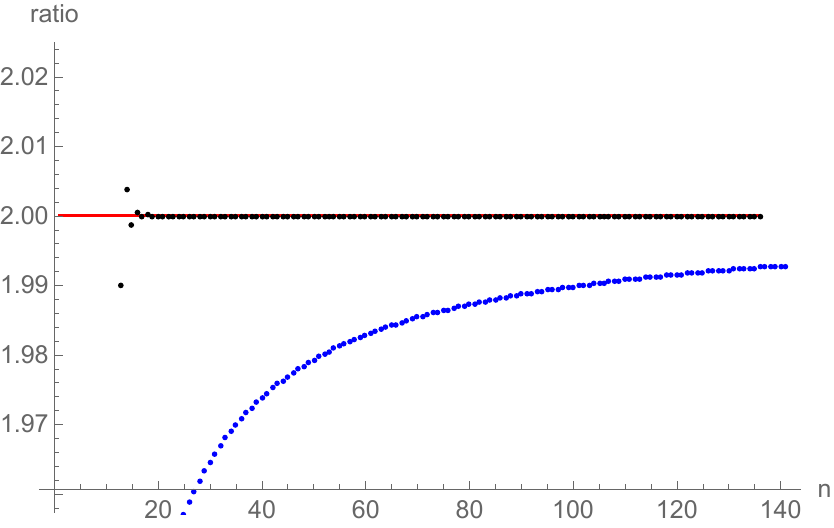}
\includegraphics[scale=0.6]{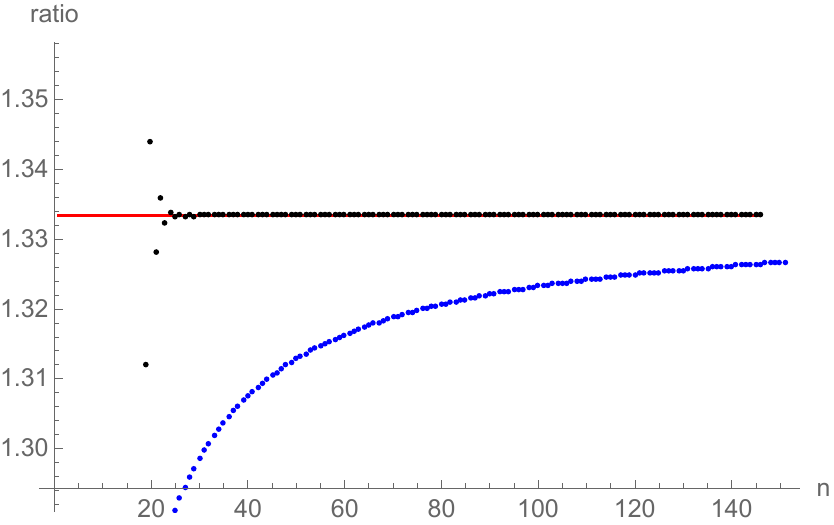}
\caption{The ratio on the left-hand-side of \eqref{eq:c-ratio} for the cusp [left] with $a=\frac{1}{4}$, and the tilted cusp [right] with $a=\frac{1}{8}$, determining the constant $A$ to be as in \eqref{eq:a-value}. The blue dots show the raw ratio \eqref{eq:c-ratio}, and the black dots show the 5th order Richardson acceleration, converging rapidly to $1/(1-2a)$.}
\label{fig:as}
\end{figure}
The large order growth parameters $A$, $\beta$, $\mathcal S$, and $b$ all potentially depend on the tilt parameter $a$.
The parameter $A$ tells us the location of the leading Borel singularity. The parameter $\beta$ determines the exponent of this leading singularity, and $\mathcal S$ tells us the strength of this leading singularity. The parameter $b$ in the subleading power-law corrections tells us about the analytic fluctuations about the leading singularity. This information is all encoded in the strong coupling expansion coefficients $c_n(a)$ in \eqref{eq:gamma-strong}, and can be extracted numerically, by a sequence of methods of increasing precision. 
\begin{figure}[h!]
\includegraphics[scale=0.6]{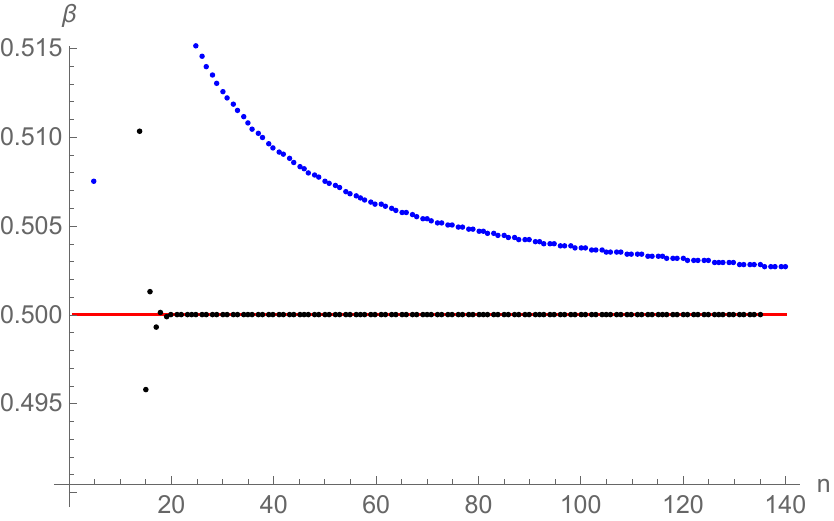}
\includegraphics[scale=0.6]{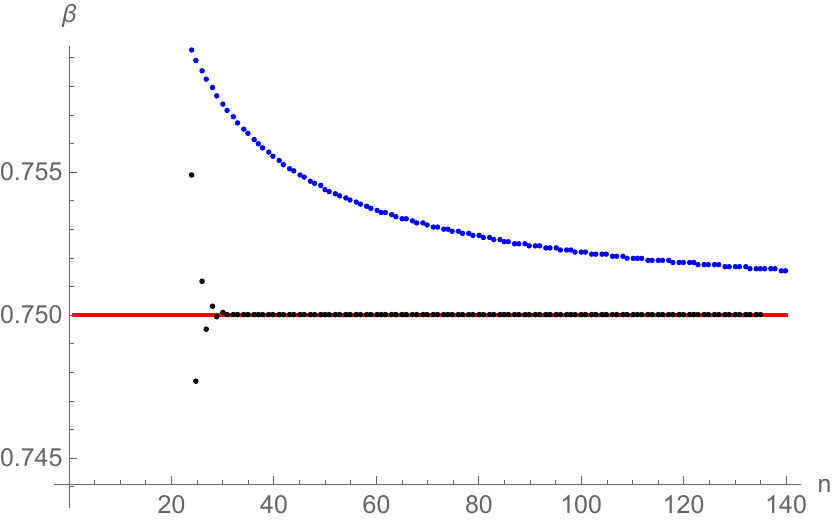}
\caption{The ratio on the left-hand-side of 
\eqref{eq:beta-ratio} for the cusp [left] with $a=\frac{1}{4}$, and the tilted cusp [right] with $a=\frac{1}{8}$. The blue dots denote the raw combination, and the black dots show the 5th order Richardson acceleration, converging to \eqref{eq:beta-value}.}
\label{fig:betas}
\end{figure}
A first rough guide can be obtained by ratio tests, enhanced with Richardson acceleration \cite{bender}. The ratio of successive terms $c_{n+1}/c_n$ scales like $n$, so we can use this fact to isolate the parameter $A$:
\begin{eqnarray}
    \frac{c_{n+1}}{n\, c_n}\sim
    \frac{1}{A}-\frac{\beta }{A n}-\frac{b}{n^2} +O\left(\frac{1}{n^3}\right)
    \label{eq:c-ratio}
\end{eqnarray}
Using Richardson acceleration one finds (see Figure \ref{fig:as}) that the constant $A$ has a simple linear dependence on the tilt parameter $a$:
\begin{eqnarray}
    A=(1-2a) 
    \label{eq:a-value}
\end{eqnarray}
Given $A$, the next term in \eqref{eq:c-ratio} identifies $\beta$ as
\begin{eqnarray}
    n\left(1-\frac{A\, c_{n+1}}{n\, c_n}\right) \sim  \beta+O\left(\frac{1}{n}\right)
    \label{eq:beta-ratio}
\end{eqnarray}
We find that $\beta$ is the same linear function of the tilt parameter $a$ (see Figure \ref{fig:betas}):
\begin{eqnarray}
    \beta=(1-2a)
    \label{eq:beta-value}
\end{eqnarray}
Given $A$ and $\beta$, the next term in \eqref{eq:c-ratio} determines the parameter $b$ in the subleading power-law correction in \eqref{eq:cn}:
\begin{figure}[h!]
\includegraphics[scale=0.6]{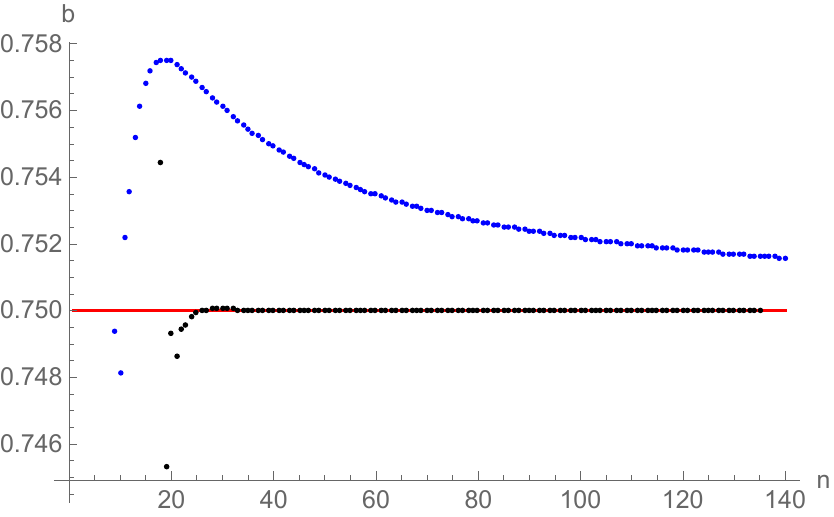}
\includegraphics[scale=0.6]{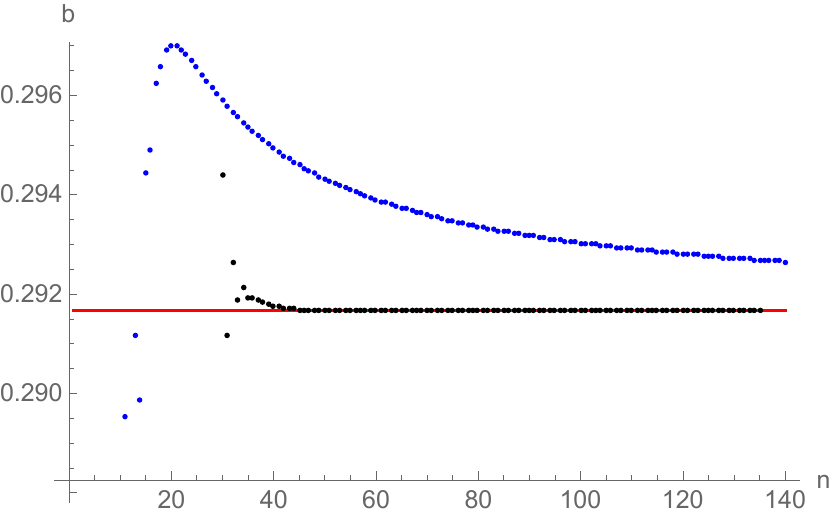}
\caption{The combination on the left-hand-side of \eqref{eq:b-ratio} determines the parameter $b$ to be as in \eqref{eq:b-value}. The blue dots denote the raw combination, and the black dots show the 5th order Richardson acceleration, converging to \eqref{eq:b-value}.}
\label{fig:bs}
\end{figure}
\begin{eqnarray}
    \frac{n}{A}\left(n\left(1-\frac{A\, c_{n+1}}{n\, c_n}\right) -\beta\right)\sim   b +O\left(\frac{1}{n}\right)
    \label{eq:b-ratio}
\end{eqnarray}
The dependence of $b$ on the tilt parameter $a$ is found to be (see Figure \ref{fig:bs}):
\begin{eqnarray}
    b=\frac{2a(1-a)}{(1-2a)}
    \label{eq:b-value}
\end{eqnarray}
So, elementary analysis of the strong coupling expansion coefficients reveals that the leading large order growth is:
\begin{eqnarray}
c_n(a) \sim {\mathcal S}_a \frac{\Gamma(n-(1-2a))}{(1-2a)^n}\left[1+\frac{2a(1-a)}{(n-(1-2a)-1)}+\dots\right]+\dots
\label{eq:large-order-c}
\end{eqnarray}
If at any stage in this succession of ratio tests the conjectured $a$ dependence were incorrect, the next test would fail dramatically.
\begin{figure}[h!]
\includegraphics[scale=0.6]{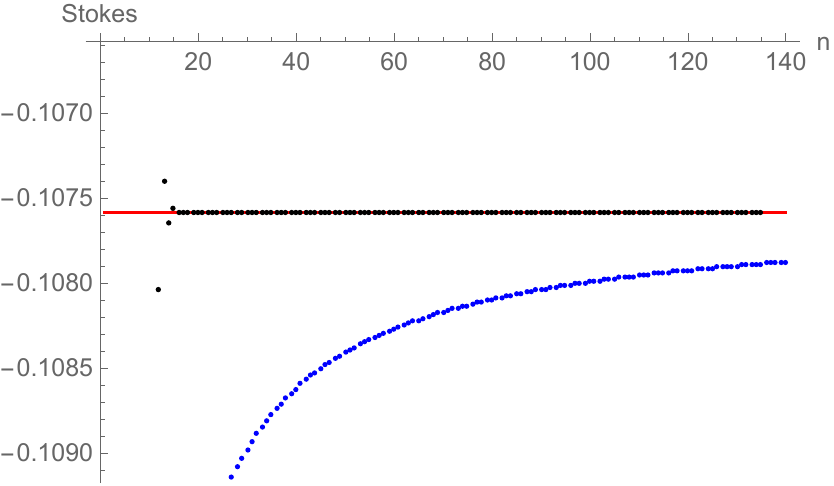}
\includegraphics[scale=0.6]{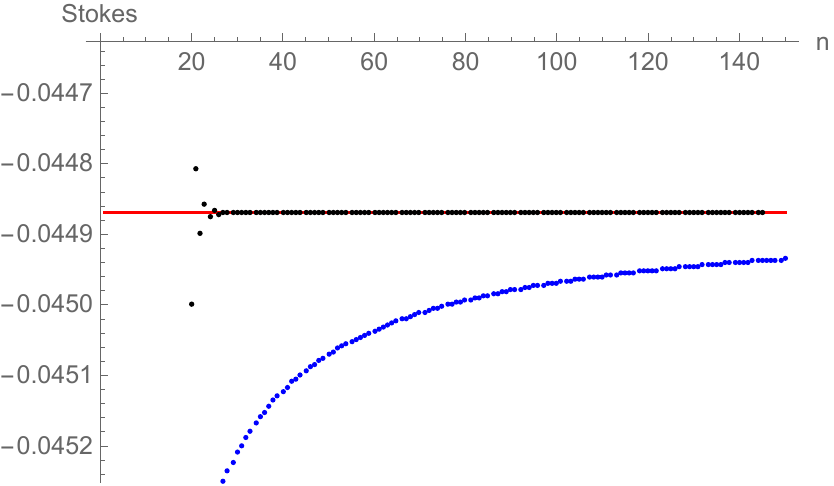}
\caption{Ratio tests using expression \eqref{eq:large-order-c} to determine the Stokes constant $\mathcal S_a$ for $a=\frac{1}{4}$ [left] and $a=\frac{1}{8}$ [right]. The blue dots show the raw ratio, as a function of the perturbative order $n$, and the black dots show the 5th order Richardson acceleration, converging rapidly  to the values indicated by the red line, from \eqref{eq:sa14}, \eqref{eq:sa18} and from the analytic expression \eqref{eq:stokes-formula}.}
\label{fig:stokes}
\end{figure}
The overall constant factor ${\mathcal S}_a$ (the Stokes constant) in Equation \eqref{eq:large-order-c} is generally the most difficult thing to determine from the large-order behavior. Given the information in \eqref{eq:large-order-c}, ratio tests produce accurate estimates. Conjectured natural expressions can then be confirmed to high precision.
The dependence of the Stokes constant ${\mathcal S}_a$ on $a$ is not a simple rational function. For some special $a$ values we find: 
\begin{eqnarray}
a=\frac{1}{4} \quad &:& \quad {\mathcal S}_{1/4}= -\frac{1}{4\pi} \frac{\Gamma\left(\frac{3}{4}\right)}{\Gamma\left(\frac{5}{4}\right)}
\label{eq:sa14}\\
a=\frac{1}{3} \quad &:& \quad {\mathcal S}_{1/3}= -\frac{1}{2^{5/3}\, \pi} \frac{\Gamma\left(\frac{2}{3}\right)}{\Gamma\left(\frac{4}{3}\right)}
\label{eq:sa13}\\
a=\frac{1}{6} \quad &:& \quad {\mathcal S}_{1/6}= -\frac{1}{6\pi} \frac{\Gamma\left(\frac{5}{6}\right)}{\Gamma\left(\frac{7}{6}\right)}
\label{eq:sa16}\\
a=\frac{1}{8} \quad &:& \quad {\mathcal S}_{1/8}=  -0.0448694168776214163...  
\label{eq:sa18}\\
a=\frac{1}{10} \quad &:& \quad {\mathcal S}_{1/10}=-0.0339838733... 
\label{eq:sa110}
\end{eqnarray}
The $a=\frac{1}{4}$ Stokes constant was known previously in \cite{Basso:2007wd,Basso:2009gh,Aniceto:2015rua,Dorigoni:2015dha}. 
The values above suggest a factor $\frac{\Gamma(1-a)}{\Gamma(1+a)}$ in $\mathcal S_a$, and this is certainly natural from the work of Basso et al \cite{Basso:2007wd,Basso:2009gh}. The fact that the strong coupling expansion truncates for the octagon ($a=0$) also implies that the Stokes constant should vanish when $a=0$. The limit $a\to \frac{1}{2}$ is also special, and is best studied in a double-scaling limit \cite{unpub,Bajnok:2024bqr}.

To fix the $a$-dependence of the Stokes constant, recall that the large order behavior \eqref{eq:large-order-c} refers to the coefficients of the strong coupling expansion \eqref{eq:gamma-strong} in terms of the shifted coupling parameter $\xi$ defined in \eqref{eq:xi}. Converting back to an expansion in terms of $(4\pi g)$, the large order behavior acquires an extra factor 
\begin{eqnarray}
    \exp\left[\frac{s_1(a)}{2} (1-2a)\right]
    \label{eq:factor}
\end{eqnarray}
where (here $\gamma$ is the Euler gamma constant and $\psi$ is the digamma function):
\begin{eqnarray}
    s_1(a)=-2\gamma-\psi\left(\frac{1}{2}+a\right)-\psi\left(\frac{1}{2}-a\right)
    \label{eq:s1a}
\end{eqnarray}
With this conversion factor included we find a simple expression for the Stokes constant:
\begin{eqnarray}
    \mathcal S_a= \exp\left[-\frac{s_1(a)}{2} (1-2a)\right]\times\left(-\frac{\sin(a \pi)}{\pi}\frac{\Gamma(1-a)}{\Gamma(1+a)} \right)
    \label{eq:stokes-formula}
\end{eqnarray}
This agrees \cite{grisha} with the values in \eqref{eq:sa14}-\eqref{eq:sa110} to all known digits. The second factor is the result from the analytic approach in \cite{Basso:2009gh,Bajnok:2024epf,Bajnok:2024ymr} in terms of the coupling $g$ instead of the shifted coupling $\xi$.

The final dots in the large-order growth expression \eqref{eq:large-order-c} denote  possible exponentially suppressed corrections, due to more distant Borel singularities. These are much more difficult to extract using simple ratios tests, as they are overwhelmed by the power-law corrections in \eqref{eq:large-order-c}. Indeed, we show in the next Section that there exist new more distant singularities, and moreover these are all repeated in integer multiples because the underlying problem is nonlinear. To learn about this rich Borel singularity structure we need better tools than just ratio tests and Richardson acceleration. So we turn to the more powerful Borel methods. Note that the Borel analysis is based on exactly the same input information (i.e., the strong-coupling expansion coefficients), but is able to extract more refined information.

\subsection{Borel Analysis of the Strong Coupling Expansion}
\label{sec:borel}

The accuracy of the extrapolation of the asymptotic strong coupling expansion is determined by the accuracy of the analytic continuation of the Borel transform, especially near its singularities. The easiest approach is to use Pad\'e to continue the Borel transform, but it is significantly more accurate to first make a conformal map. Even more refined information can be obtained by the method of {\it singularity elimination} \cite{Costin:2020pcj}, which effectively removes a chosen singularity, thereby permitting more precise analysis of that region. See Section \ref{sec:sing}. We take as input data a list of approximately 150 strong coupling coefficients, with 300 digit precision, and then define the Borel transform by dividing the coefficients by $n!$
\begin{eqnarray}
    B_a(\zeta):=\sum_{n=0}^\infty \frac{c_n(a)}{n!} \, \zeta^n
    \label{eq:borel}
\end{eqnarray}
The formal divergent strong coupling expansion \eqref{eq:gamma-strong}-\eqref{eq:strong-a} is recovered by the Laplace transform
\begin{eqnarray}
    \Gamma_a(\xi) =\frac{2 a\, \xi^2}{\pi \sin(2\pi a)} \int_0^\infty d\zeta\, e^{-\zeta\, \xi}\, B_a(\zeta)
    \label{eq:laplace}
\end{eqnarray}
Having dividing out the factorial growth of the strong coupling  coefficients $c_n(a)$, the Borel transform $B_a(\zeta)$ has a finite radius of convergence, which means that it has at least one singularity away from the origin in the finite complex Borel $\zeta$ plane. In order to obtain an accurate extrapolation in the physical $\xi$ plane, we need to extract as much high-precision information as possible about the behavior of the Borel transform function in the neighborhood of the singularities, especially the leading singularity. 

\subsubsection{Pad\'e-Borel Analysis}
\label{sec:pb} 

\begin{figure}[h!]
\centerline{ \includegraphics[scale=.65]{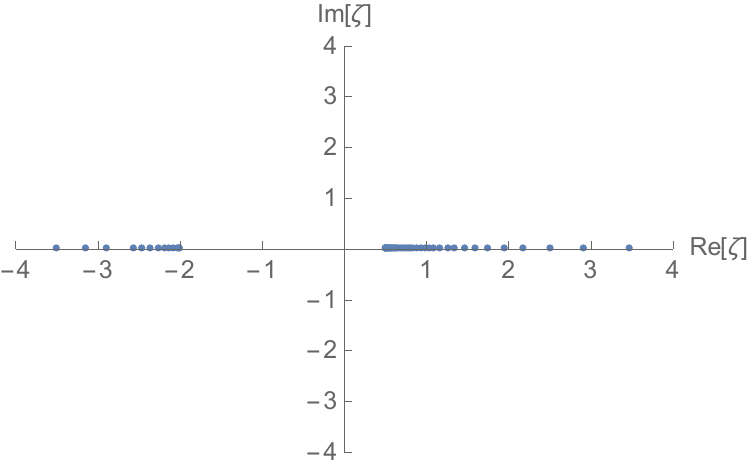}
\quad \includegraphics[scale=.65]{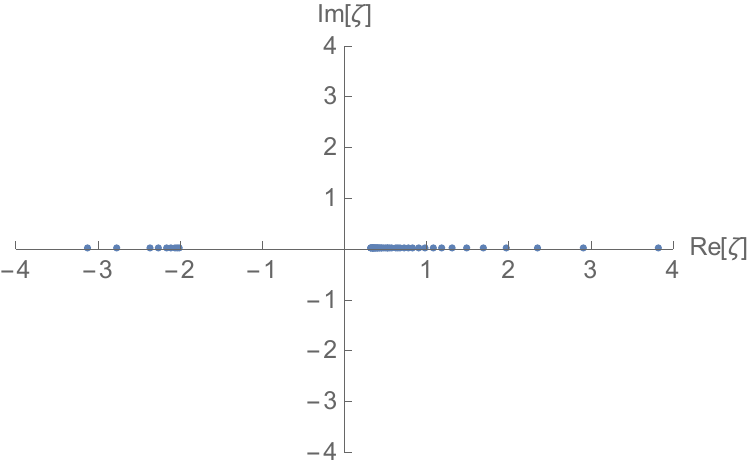}}
\centerline{\includegraphics[scale=.65]{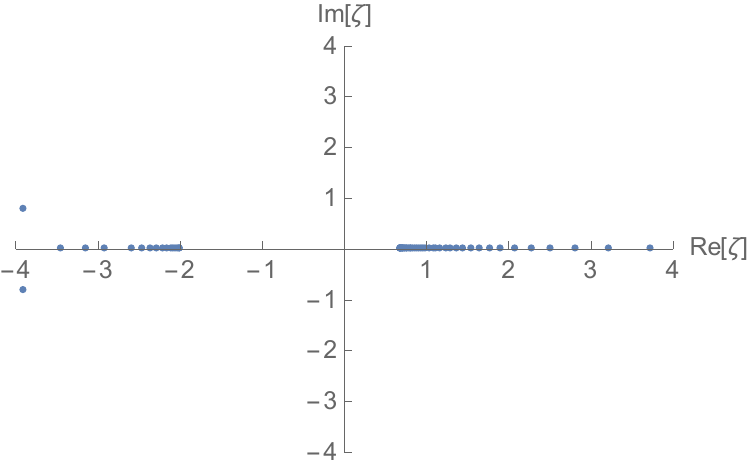}
\quad 
\includegraphics[scale=.65]{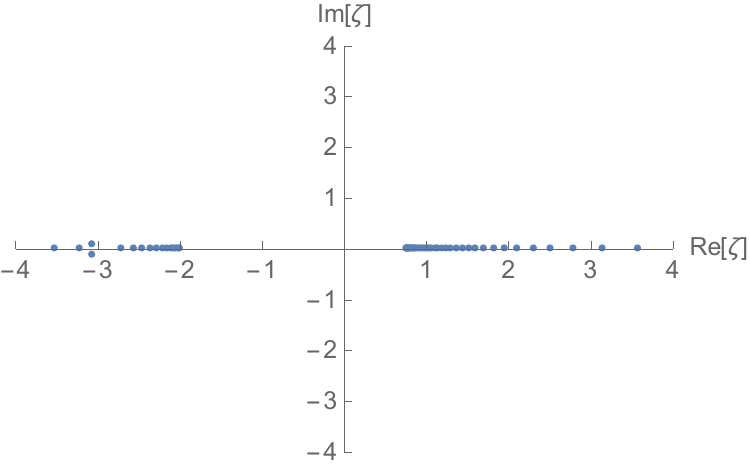}}
\caption{The Pad\'e poles of the Borel transform, for $a=\frac{1}{4}$ [top left], $a=\frac{1}{3}$ [top right], $a=\frac{1}{6}$ [bottom left], $a=\frac{1}{8}$ [bottom right]. We see a leading branch point singularity at $\zeta=(1-2a)$, and another at $\zeta=-2$, independent of $a$.  }
\label{fig:pade-poles}
\end{figure}

The numerical goal is to learn as much precise information as possible about the singularities of the Borel transform, as these correspond to the non-perturbative physics.
But since the strong coupling expansion is truncated at a finite order, this means we are trying to probe the singularities of the Borel function given only a {\it polynomial approximation} to the Borel function. The simplest way to obtain a rough idea of the Borel singularity structure is to make a Pad\'e approximation of the truncated Borel transform. However, Pad\'e produces an analytic continuation of the truncated Borel transform function that is a ratio of polynomials, so it can only have pole singularities. But we use the fact that Pad\'e generically represents a branch point as the accumulation point of an arc of interlacing poles and zeros, based on the electrostatic interpretation of Pad\'e \cite{Stahl,aptekarev,Costin:2021bay}. For the tilted cusp this is illustrated in Figure \ref{fig:pade-poles}. For a selected set of tilt parameters, $a\in \{ \frac{1}{4}, \frac{1}{3},\frac{1}{6}, \frac{1}{8}\}$, we make diagonal Pad\'e approximants of the truncated Borel transform and plot the resulting Pad\'e poles. We observe that there is a leading branch point singularity on the positive real axis of the Borel plane, whose location is a simple function of $a$:
\begin{eqnarray}
    \zeta_{\rm leading}=(1-2a)
    \label{eq:zeta-leading}
\end{eqnarray}
This confirms the ratio test result \eqref{eq:a-value} that the leading Borel singularity is at $\zeta_{\rm leading}=A=(1-2a)$.

More interestingly, we observe in Figure \ref{fig:pade-poles} that there is an additional Borel singularity at $\zeta=-2$, independent of the value of the tilt parameter $a$, and which is more distant (recall $0\le a <1/2$):
\begin{eqnarray}
    \zeta_{\rm negative}=-2
    \label{eq:zeta-negative}
\end{eqnarray}
In fact, this is not the whole story. Since this is a nonlinear problem, we should expect that these Borel singularities are repeated in integer multiples of $(1-2a)$, and also at integer multiples of $-2$. These are not visible in the Pad\'e pole plots in Figure \ref{fig:pade-poles}, because these repeated singularities are hidden among the accumulating poles that Pad\'e produces when representing a branch cut. Fortunately there is a simple way to resolve the existence of these repeated singularities, by making a suitable conformal map, {\it before} making the Pad\'e approximation.

\subsubsection{Pad\'e-Conformal-Borel Analysis}
\label{sec:pcb}

The procedure of the Pad\'e-Conformal-Borel analysis is a simple generalization of the Pad\'e-Borel method described in the previous Section, but it has two distinct advantages \cite{Costin:2020hwg}. First, it reveals the existence of higher singularities that might be hidden among the Pad\'e poles that appear to be representing a branch cut. Second, it leads to a significantly more accurate analytic continuation of the truncated Borel transform function, which leads to a significantly more accurate analytic continuation of the physical quantity, the tilted-cusp anomalous dimension $\Gamma_a(\xi)$. The Pad\'e-Conformal-Borel procedure mirrors the Pad\'e-conformal extrapolation procedure for the convergent weak coupling expansion in Section \ref{sec:weak-exrapolation}, but now applied to the Borel transform, not the cusp function.
\begin{enumerate}
\item First, go to the Borel plane. It is significantly more accurate to do the Pad\'e analysis in the Borel plane than in the original physical variable \cite{Costin:2020hwg}.

\item Make a conformal map from the Borel variable $\zeta$ to a new variable, $z$, which maps the cut Borel plane to the interior of the unit disk. In practice, one does not know the entire singularity structure in the Borel plane, but it is already an excellent approximation to use a conformal map based on the known {\it leading} Borel singularities. For example, given the information provided by the simple Pad\'e-Borel analysis of the tilted cusp (recall Figure \ref{fig:pade-poles}), we base our conformal map on the approximation that there is a cut along the positive real axis $\zeta\in [1-2a, \infty)$, and also along the negative axis $\zeta\in (-\infty, -2]$. This two-cut Borel plane is mapped to the interior of the unit disk $|z|<1$ by the following conformal map
    \begin{eqnarray}
\zeta=\frac{8(1-2a)z}{2(1+z)^2+(1-2a)(1-z)^2} 
\quad \longleftrightarrow \quad 
z=\frac{\sqrt{(1-2a)(2+\zeta)}-\sqrt{2(1-2a-\zeta)}}{\sqrt{(1-2a)(2+\zeta)}+\sqrt{2(1-2a-\zeta)}}
\label{eq:cmap}
\end{eqnarray}
The upper/lower edges of the cuts are mapped to sections of the upper/lower semicircular {\it boundaries} of the unit disk, and the region away from the two cuts in the $\zeta$ plane is mapped to the {\it interior} of the unit disk.
\item Re-expand the truncated Borel transform in $z$ to the same order (this is provably optimal \cite{Costin:2020pcj}).
\item Now make a Pad\'e approximation to this mapped Borel transform, in terms of the variable $z$.
\item Map back to the Borel $\zeta$ plane using the inverse conformal map.
\end{enumerate}

\noindent{\bf Remarks:}
\begin{itemize}
    \item The chosen conformal map only depends on the {\it location} of the branch points, not on their associated exponents. 
    \item If the conformal map happens to be based on {\it all} the actual $\zeta$ singularities of the Borel transform, then by construction there cannot be any singularities inside the unit disk in the $z$ plane. 
    \item If there are other (unknown in advance) singularities that were hidden underneath the Pad\'e-Borel poles, and which were therefore not incorporated in the chosen conformal map, then the new Pad\'e poles in the $z$ plane will accumulate to these singularities,  which must lie on the unit circle. And since the actual singularities lie on a higher Riemann sheet, the accumulating arc of Pad\'e poles will approach the unit disk from the exterior, not from the interior. This is exactly what happens for the tilted cusp. See Section \ref{sec:pcb-results} and Figure \ref{fig:z-poles-1-eighth} below.
    \item Given only a finite amount of information, in the sense of being given only a finite order of truncation (due to having only a finite number of terms in the original expansion), and/or in the sense of having only finite precision for the coefficients, then some Pad\'e poles may leak into the interior of the unit disk in the mapped $z$ plane. These can be filtered by analyzing sequences of near-diagonal Pad\'e approximants.
    \item Conformal maps are known for many configurations of few branch points, and for situations where the branch point locations are symmetrically distributed in the complex plane \cite{kober}. For more complicated singularity configurations, it is still a significant numerical improvement to make a conformal map based on only the leading few branch point locations.
\end{itemize}

\subsubsection{Results of Pad\'e-Conformal-Borel Analysis}
\label{sec:pcb-results}

\begin{figure}[htb]
\centerline{
\includegraphics[scale=.75]{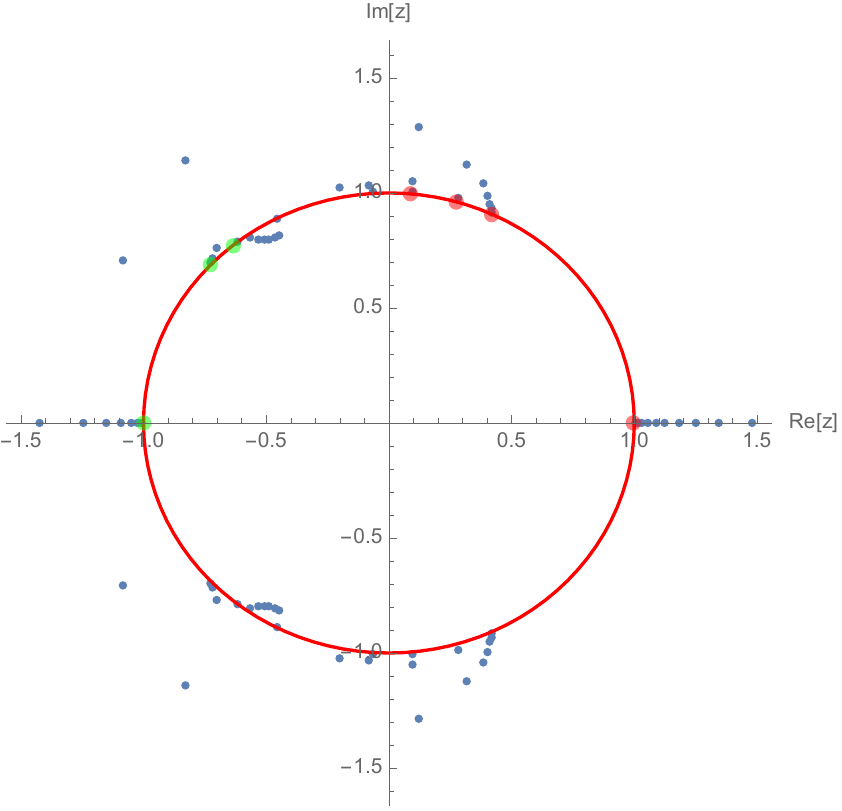}}
\caption{The Pad\'e poles (blue dots) in terms of the conformally mapped variable $z$ defined in \eqref{eq:cmap}, for $a=\frac{1}{8}$. Poles appear outside the unit disk, accumulating to points on the unit circle, which are the $z$-plane images of true Borel singularities lying on the cuts along the real $\zeta$ axis in the Borel plane. The singularities for $Re(\zeta)>0$ are shown in red, and those for $Re(\zeta)<0$ are shown in green. The leading singularity at $\zeta=(1-2a)=\frac{3}{4}$ maps to $z=+1$, while the singularity at $\zeta=-2$ maps to $z=-1$. Higher Borel singularities map to points on the unit circle, also marked in red and green. The next most dominant singularity is at $\zeta=(1+2a)=\frac{5}{4}$, which maps to $z=0.418 + 0.908 i$; followed by $\zeta=2(1-2a)=\frac{3}{2}$, which maps to $z=0.273 + 0.962 i$;  followed by $\zeta=(1-2a)+(1+2a)=2 $, which maps to $z=0.091 + 0.996 i$. 
The Borel singularity at $\zeta=-4$ maps to $z=-0.727 + 0.686 i$, and $\zeta=-6$ maps to $z=-0.636 + 0.771 i$. 
The small arc of poles inside the disc near $z=-0.591 + 0.807 i$, the image of $\zeta=-8$, indicates a breakdown due to the limited number of terms and the limited precision used.
The point at infinity in the Borel plane maps to $z=-0.455 + 0.891 i$.}
\label{fig:z-poles-1-eighth}
\end{figure}

We use the conformal map \eqref{eq:cmap} based on two cuts along the real axis of the Borel plane, emanating from $-2$, and from $(1-2a)$. Recall Figure \ref{fig:pade-poles}. The resulting Pad\'e poles in the conformally mapped $z$ plane are shown in Figure \ref{fig:z-poles-1-eighth} for a chosen tilt parameter $a=\frac{1}{8}$. This analysis can be done for any $a$, but this value is convenient for a reason that will become clear below.

The first thing we learn from this Pad\'e-Conformal-Borel analysis is that there is a new singularity on the positive real Borel axis at $\zeta=(1+2a)$, which corresponds to a pair of complex conjugate points on the unit circle in the $z$ plane, and  whose location is $a$-dependent: 
\begin{eqnarray}
\zeta_{\rm new} =(1+2a)
  \qquad \longleftrightarrow \qquad 
z_{\rm new}=\frac{\sqrt{(1-2a)(3+2a)}\pm i \sqrt{8a}}{\sqrt{(1-2a)(3+2a)}\mp i \sqrt{8a}}
\label{eq:zeta-new}
\end{eqnarray}
This new Borel singularity is more distant (and therefore sub-dominant) than the leading one at $\zeta_{\rm leading}=(1-2a)$, but closer to the origin (and therefore more dominant) than the one at $\zeta_{\rm negative}=-2$. This singularity is not visible in Figure \ref{fig:pade-poles}, because it is hidden among the associated Pad\'e poles that represent the first cut.

The next thing we learn is that the Borel singularities at $\zeta_{\rm leading}=(1-2a)$, at $\zeta_{\rm new}=(1+2a)$, and at $\zeta_{\rm negative}=-2$ are repeated in integer multiples. The numerical results of the Pad\'e-Conformal-Borel analysis suggest that there is a doubly-infinite tower of Borel singularities along the positive real Borel axis, and an infinite tower of Borel singularities along the negative real Borel axis:
\begin{eqnarray}
\zeta^{(p, q)}_{\rm positive}:=p(1-2a)+q(1+2a) \quad &\longleftrightarrow& \quad
z^{(p, q)}_{\rm positive}=
   \frac{1\pm i \sqrt{2} \sqrt{\frac{(1-2 a) (p-1)+(1+2 a) q}{(1-2 a) ((1-2 a) p+(1+2 a)
   q+2)}}}{1\mp i \sqrt{2} \sqrt{\frac{(1-2 a) (p-1)+(1+2 a) q}{(1-2 a) ((1-2 a) p+(1+2 a)
   q+2)}}}
   \\
 \zeta^{(k)}_{\rm negative}:=-2 \, k   \quad &\longleftrightarrow& \quad
z^{(k)}_{\rm negative}=  \frac{\sqrt{(1-2a)(1-k)} +\sqrt{1-2a+2k}}{\sqrt{(1-2a)(1-k)} -\sqrt{1-2a+2k}}
   \end{eqnarray}
   Here $p$ and $q$ are integers ("instanton numbers" or "transseries indices"), starting with $(p, q)=(1,0)$ for the leading Borel singularity, and $k\geq 1$ with $k=1$ denoting the leading singularity at $\zeta=-2$ on the negative real axis.
The point at infinity in the Borel $\zeta$ plane maps to a complex conjugate pair of points on the unit circle in the $z$ plane, whose location depends on the tilt parameter $a$:
\begin{eqnarray}
    \zeta=\infty  \quad &\longleftrightarrow& \quad z_{\rm infinity} =
    \frac{\sqrt{1-2a}\pm i \sqrt{2}}{\sqrt{1-2a}\mp i \sqrt{2}}
\end{eqnarray}
The existence of other Borel singularities beyond the leading one at $\zeta_{\rm leading}=(1-2a)$ implies that there will also be exponential corrections to the large-order behavior in (\ref{eq:large-order-c}). In the ratio test analysis of the large-order growth in Section \ref{sec:growth} these exponential corrections are completely swamped by the power-law corrections, so they are effectively inaccessible. However, the Pad\'e-Conformal-Borel method can resolve them.

To illustrate this structure most clearly, we have chosen a value of the tilt parameter, $a=\frac{1}{8}$, because the first new singularity, at $\zeta=1+2\times \frac{1}{8}=\frac{5}{4}$ is closer to the origin than two times the leading singularity, at $\zeta= 2\times \frac{3}{4}=\frac{3}{2}$.

Beyond the existence of the new Borel singularity \eqref{eq:zeta-new}, we see a few integer repetitions of the singularities at $\zeta_{\rm leading}$, $\zeta_{\rm new}$, and $\zeta_{\rm negative}$. Interestingly, we see a singularity at $\zeta_{\rm leading}+\zeta_{\rm new}=+2$. 
See Figure \ref{fig:z-poles-1-eighth}. For example, for $a=\frac{1}{8}$ we have (in order of dominance, Figure \ref{fig:z-poles-1-eighth}):
\begin{eqnarray}
\zeta_{\rm positive}^{(1,0)}=(1-2a)=\frac{3}{4} 
\quad &\longrightarrow& \quad  z_{\rm positive}^{(1,0)} =+1
\nonumber\\
\zeta_{\rm positive}^{(0,1)}=(1+2a)=\frac{5}{4} \quad &\longrightarrow& \quad  z_{\rm positive}^{(0,1)} =0.418 + 0.908 i
\nonumber\\
\zeta_{\rm positive}^{(2,0)}=2(1-2a)=\frac{3}{2} \quad &\longrightarrow& \quad  z_{\rm positive}^{(2,0)} =0.273 + 0.962 i
\nonumber\\
\zeta_{\rm positive}^{(1,1)}=(1-2a)+(1+2a)=2 \quad &\longrightarrow& \quad  z_{\rm positive}^{(1,1)} =0.091 + 0.996 i
\nonumber\\
\zeta_{\rm negative}^{(1)}=-2 \quad &\longrightarrow& \quad  z_{\rm negative}^{(1)} =-1
\nonumber\\
\zeta_{\rm negative}^{(2)}=-4 \quad &\longrightarrow& \quad  z_{\rm negative}^{(2)} =-.727+0.686 i
\nonumber\\
\zeta_{\rm positive}^{(0,2)}=2(1+2a)=\frac{5}{2} \quad &\longrightarrow& \quad  z_{\rm positive}^{(0,2)} =-0.0182 + 0.999 i
\end{eqnarray}
With more terms, and higher precision, it would be possible to resolve further singularities. The main point is that this is strong numerical evidence for the existence of the second independent Borel singularity, and also integer multiple repetitions of the three fundamental Borel singularities: $\zeta_{\rm leading}$, $\zeta_{\rm new}$, and $\zeta_{\rm negative}$. And recall that all this non-perturbative information has been decoded from the perturbative strong coupling expansion, without any reference to the equations that generated the strong coupling expansion.

\subsubsection{Singularity Elimination}
\label{sec:sing}

{\it Singularity elimination} is a powerful method to probe higher Borel singularities \cite{Costin:2020pcj}. The application of a linear operator followed by a suitable conformal map completely removes a chosen singularity, thereby enabling access to the fluctuations near the location of the removed singularity, and also access to higher Riemann sheets. Here we illustrate the procedure by removing the leading Borel singularity, leading to significantly higher precision for the Stokes constant compared to the ratio tests in Section \ref{sec:growth}.

\begin{figure}[h!]
\centerline{
\includegraphics[scale=.75]{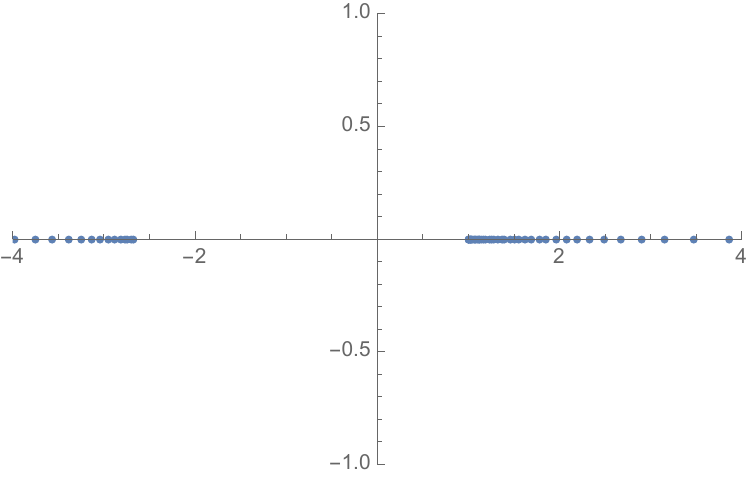}}
\caption{Rescaled Pad\'e poles in the $\zeta$ plane, before the conformal map, normalized to have the leading singularity at $\zeta=\frac{4}{3}\times \frac{3}{4} =+1$, and the one on the negative axis at $\zeta=\frac{4}{3}\times (-2)=-\frac{8}{3}$. 
}
\label{fig:a18-pb-poles}
\end{figure}

To simplify the analysis we first rescale the Borel variable, dividing by $(1-2a)$, in order to place the leading singularity at $\zeta=1$. With this rescaling of the Borel plane, the singularity $\zeta_{\rm negative}$ now appears at $\frac{-2}{1-2a}$, and $\zeta_{\rm new}$ now appears at $\frac{1+2a}{1-2a}$: see Figure \ref{fig:a18-pb-poles} for the rescaled Borel $\zeta$ plane for $a=\frac{1}{8}$. This singularity has exponent $\beta=1-2\times \frac{1}{8}=\frac{3}{4}$.

\begin{figure}[htb]
\centerline{
\includegraphics[scale=.75]{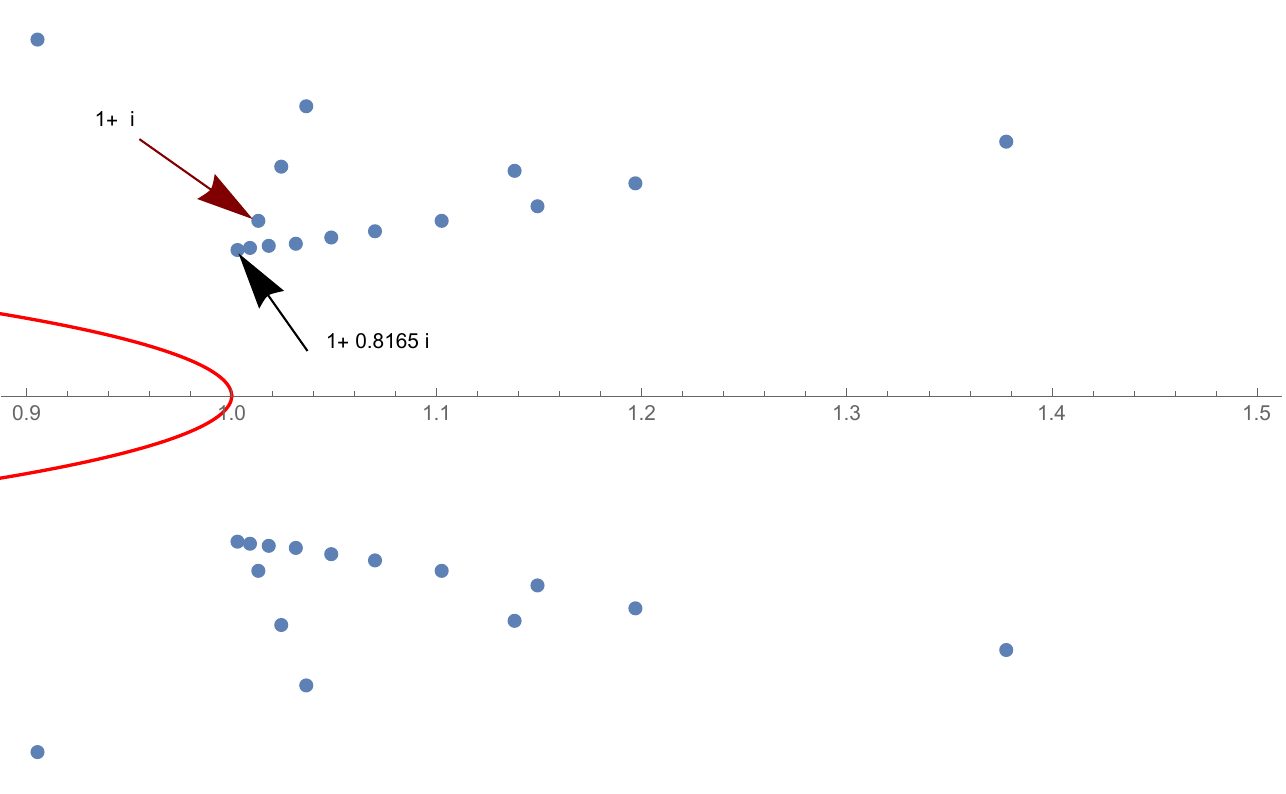}}
\caption{The poles in the new $z$ plane after the singularity elimination map: $\zeta=2z-z^2$. Note that the leading singularity at $z=1$ has been completely  removed. Now the leading singularities in the right hand plane are at $z=1-\sqrt{1-\zeta}$ with $\zeta=\frac{5}{3}$, namely at $z=1\pm 0.8165 i$ (marked with the black arrow). We can also see a singularity at $z=1-\sqrt{1-\zeta}$ with $\zeta=2$, namely at $z=1\pm  i$ (marked with the red arrow).}
\label{fig:a18-elim-poles-closeup}
\end{figure}

\begin{enumerate}

\item
The first step is to convert the exponent $\beta=1-2 a=\frac{3}{4}$ to a new exponent $\frac{1}{2}$, by a fractional derivative. See \cite{Costin:2020pcj}, equation (30). Here the original exponent $\beta=1-2\times \frac{1}{8}=\frac{3}{4}$, so to achieve a new exponent equal to $\frac{1}{2}$, we need to choose the fractional derivative parameter to be $\gamma$ such that
\begin{eqnarray}
\beta+\gamma+1=\frac{1}{2} \qquad \Rightarrow \qquad \gamma=-\frac{5}{4}
\end{eqnarray}

\item
This defines a new series: see eq 30 in \cite{Costin:2020pcj}:
\begin{eqnarray}
\tilde{B}(\zeta)=\sum_{n=0}^{150} \frac{\Gamma\left(1+\gamma\right)\Gamma\left(n+1\right)}{\Gamma\left(n+2+\gamma\right)} a_n \zeta^{n+1}
\end{eqnarray}

\item
Now we re-expand this modified Borel transform as $\tilde{B}(\zeta=2z-z^2)$ in powers of the mapped variable $z$, and make a Pad\'e approximant in $z$. Since the new exponent is $\frac{1}{2}$, this conformal map $\zeta = 2z-z^2$ removes the square root singularity: $\sqrt{1-(2z-z^2)}=1-z$. There is no longer any branch point at $z=1$ (which is the conformal map image of $\zeta=1$). The resulting $z$ poles are plotted in  Figure \ref{fig:a18-elim-poles-closeup}.
\end{enumerate}

After the elimination procedure, we see that the singularity at $z=1$ (i.e. at $\zeta=1$) has been eliminated, enabling us to see more clearly the next 2 singularities, indicated in the figure. 
Furthermore, expanding $\tilde{B}(\zeta=2z-z^2)$ near the eliminated singularity we find
\begin{eqnarray}
    \tilde{B}(2z-z^2) = -3.44359985603875099186782189 + 
 0.779649032524983941777425693 (z-1)+ O((z-1)^2)
    \label{eq:singz1}
\end{eqnarray}
Note that after reversing the elimination map the coefficient of $(z-1)$ maps to the coefficient of the eliminated leading singularity, and we notice that
\begin{eqnarray}
    \frac{0.779649032524983941777425693495945053}{2\sqrt{\pi} \Gamma\left(1-\frac{5}{4}\right)}= -0.0448694168776214163392465863606362417
    \label{eq:stokes2}
\end{eqnarray}
This matches all known digits of the Stokes constant for the $a=\frac{1}{8}$ tilted cusp found in \eqref{eq:sa18} using ratio test methods for the large-order growth of the strong coupling expansion coefficients.
For further comparison, the analytic Stokes constant result in \eqref{eq:stokes-formula} yields for $a=\frac{1}{8}$:
\begin{eqnarray}
    \mathcal S_{\frac{1}{8}}&=& \left\{\exp\left[-\frac{s_1(a)}{2} (1-2a)\right]\times\left(-\frac{\sin(a \pi)}{\pi}\frac{\Gamma(1-a)}{\Gamma(1+a)} \right)\right\}_{a=\frac{1}{8}} \nonumber\\
    &=& -0.0448694168776214163{\color{red} 392465863606362417}
    \label{eq:stokes-formula-a18}
\end{eqnarray}
The digits in red denote those beyond the ratio test approximation \eqref{eq:sa18}, and in agreement with the singularity elimination method result \eqref{eq:stokes2}.
We see that singularity elimination roughly doubles the number of digits of precision compared to the ratio test result. This is indicative of the dramatic increase in precision that can be obtained by singularity elimination. The singularity elimination method provides more accurate access to the expansion around the leading singularity, which in turn determines the corrections to the leading growth of the strong coupling expansion coefficients in \eqref{eq:large-order-c}.

\section{Conclusion}
\label{sec:conclusion}

We have shown that resurgent extrapolation methods are capable of decoding a significant amount of analytic information about the weak coupling and strong coupling expansions of the tilted cusp, purely from the perturbative data of these expansions, without reference to the underlying equations which generated the expansions. At weak coupling we can extrapolate deep into the strong coupling regime, and also extract the exponent of the singularity that determines the finite radius of convergence. At strong coupling the expansion is asymptotic, and the formal perturbative series contains information about the singularities of the Borel transform, which in turn encode the non-perturbative physics. From the perturbative strong coupling expansion we identify the existence of two independent Borel singularities
\begin{eqnarray}
\zeta^\mp_a=(1\mp 2a) 
\label{eq:tilted-cusp-zetas}
\end{eqnarray}
and their combination: $-(\zeta_a^+ +\zeta_a^-)=-2$. Previous resurgent analyses of the cusp anomalous dimension ($a=\frac{1}{4}$) \cite{Aniceto:2015rua,Dorigoni:2015dha} identified the existence of a leading singularity at $\zeta^-=(1-\frac{2}{4})$ and another at $\zeta=-2$ [converted to our normalization]. See Figure 1 in \cite{Aniceto:2015rua}, and Figure 2 in \cite{Dorigoni:2015dha}. Both these papers also analyzed the resurgence properties of the singularity at two times the leading singularity. Furthermore, in \cite{Dorigoni:2015dha} Dorigoni and Hatsuda observed that something unusual occurs at three times the leading Borel singularity. Now we see that this is a resonance phenomenon, because $\zeta=3\times \frac{1}{2}$ coincides with the first occurrence of the new Borel singularity $\zeta=1+2\times \frac{1}{4}$. The new Borel singularity was "hidden" by an integer multiple of the leading Borel singularity.

The physical interpretation of these Borel singularities has been elucidated by Basso and Korchemsky \cite{Basso:2009gh}, and more recently by Bajnok, Boldis and Korchemsky \cite{Bajnok:2024epf,Bajnok:2024ymr,Bajnok:2024bqr}, where analysis of the BES structure explains the existence of two different 
non-perturbative scales
\begin{eqnarray}
\Lambda_{(a, \mp)}^2\sim  g^{\pm 2a}\, \exp\left[ -(1\mp 2a) 4\pi g\right]
\label{eq:instantons}
\end{eqnarray}
where $\lambda$ is the 't Hooft coupling: $\sqrt{\lambda}=4\pi g$. 
It is interesting to note that this non-perturbative information can be extracted purely from perturbative data.

\acknowledgments
I thank Lance Dixon for discussions and correspondence, and for providing high-quality perturbative data for the tilted cusp anomalous dimension which made this analysis possible. I also thank Benjamin Basso, Daniele Dorigoni, Andy Liu, and especially Gregory Korchemsky, for discussions and correspondence. It is a sad occasion, but also an honour, to dedicate this paper to the memory of Stanley Deser.  
This work was supported by the U.S. Department of Energy, Office of Science, High Energy Physics Program under Award DE-SC0010339.
I also thank the Galileo Galilei Institute for Theoretical Physics for hospitality and the INFN for partial support during the workshop “Resurgence and Modularity in QFT and String Theory”, Spring 2024.

\end{document}